\documentclass[11pt,a4paper,onecolumn]{article}
\begin{document}
\title{\bf Entanglement of Bipartite States and Quantum Teleportation: An Introduction}

\author{\textsf{Dax Enshan Koh}}
\date{\small{Data Storage Institute, DSI Building, 5 Engineering Drive 1, \\National University of Singapore, Singapore 117608 \\ $\newline$ 16 February 2009}}

\maketitle

\begin{abstract}
There has been spectacular progress in the field of quantum information in recent decades. The development of this field highlights the importance of the role of entanglement in quantum computing, quantum teleportation and quantum cryptography. These notes serve to provide a gentle introduction to the entanglement of bipartite states.  In these notes, we introduce Bell's theorem in the form derived by Clauser, Horne, Shimony and Holt. We discuss the Schmidt decomposition and the Peres-Horodecki criterion in the entanglement of pure and mixed bipartite states. Finally, we describe a teleportation protocol as an illustration of the use of entangled states.
\end{abstract}

\section{Introduction}

The study of quantum entanglement began when Einstein, Podolsky and Rosen (EPR)\cite{1} recognised in 1935 that it is possible that two spacelike-separated quantum systems could have non-classical correlations. Unfortunately, the study of entanglement was largely ignored for about 30 years until Bell proved in 1964 that quantum mechanics is incompatible with any local hidden variable theory \cite{2}. Bell constructed the famous \textit{Bell inequality} which would hold if the universe obeyed both \textit{locality} and \textit{realism} (or, in short, \textit{local realism}), which were the assumptions that EPR thought were obviously true. Locality is the assumption that events that are space-like separated cannot have any causal links with each other, and realism is the assumption that physical properties have definite values that exist independent of measurement. Bell showed that quantum mechanics violates Bell inequality, which means that local realism and quantum mechanics cannot be simultaneously true. In 1969, Clauser, Horne, Shimony and Holt (CHSH) \cite{5} took Bell inequality one step further by proposing a form of Bell inequality, known as the \textit{CHSH inequality}, that could be tested directly by experiments. Within $3$ years, the first definitive confirmation that local realism is false was provided by the Clauser-Freedman experiment \cite{6.1}, which was carried out in 1972. Over the next few decades\footnote{For a historical account of entanglement, we refer interested readers to the book \textit{Entanglement} by Amir D. Aczel \cite{6.15}.}, several other experiments with improved designs \cite{6.2} were carried out to test Bell inequality, most of which were in favor of quantum mechanics.

About 20 years after Bell inequality, physicists and computer scientists began to realize how we could use entanglement as a resource. Among the important applications of quantum entanglement are \textit{quantum teleportation} \cite{3} (which we discuss in Section \ref{telep}), which allows for the structure of states to be transferred from one location to another without traversing the space that separates the two locations, and \textit{quantum cryptography} \cite{4}, which allows for perfectly secure communication.

In these notes, we give a detailed introduction to entanglement by first introducing Bell inequalities, in the form derived by CHSH. We then derive necessary and sufficient conditions for a bipartite pure state to be entangled using the Schmidt decomposition, and derive the Peres-Horodecki criterion which gives a necessary condition for a bipartite mixed state to be separable. Finally, we describe how any arbitrary two-level pure state can be teleported using a pure maximally entangled state, and discuss the potential applications of quantum teleportation.

\section{Density Operator}
\subsection{Bloch Sphere}
We begin by introducing a graphical representation of the density operator describing a two-level system. The most general expression for a two-level pure state, known as a \emph{qubit},  is given by
\begin{equation}
|\psi \rangle = |0\rangle \cos \theta + |1 \rangle e^{i \phi} \sin\theta
\end{equation}
up to a global phase.
In matrix representation, $|\psi \rangle$ is can be written as
\begin{equation}
|\psi \rangle = \left( \begin{array}{c} \cos \theta \\ e^{i \phi} \sin \theta \end{array}  \right)
\end{equation}

The density operator corresponding to this pure state can be expressed in terms of the Pauli matrices, as follows.

\begin{eqnarray}
\rho &=& |\psi \rangle \langle \psi| \nonumber\\ &=&\left( \begin{array}{c} \cos \theta \\ e^{i \phi} \sin \theta \end{array}  \right) \left( \begin{array}{cc} \cos \theta & e^{-i \phi} \sin \theta \end{array}  \right) \nonumber\\&=&  \frac{1}{2} \left( \begin{array}{cc} 1+ \cos 2\theta & \sin 2\theta (\cos \phi - i \sin\phi) \\  \sin 2\theta (\cos \phi + i \sin \phi) & 1- \cos 2\theta \end{array} \right)\nonumber\\&=&\frac{1}{2} (I + \sigma_x \sin2\theta \cos \phi  +\sigma_y   \sin 2\theta \sin \phi + \sigma_z \cos2\theta ) \nonumber\\ &=& \frac{1}{2} (I+\hat{n} \cdot \vec{\sigma})
\end{eqnarray}
where $\sigma_x=\left( \begin{array}{cc} 0 & 1 \\ 1 &0 \end{array} \right)$, $\sigma_y=\left( \begin{array}{cc} 0 & -i \\ i &0 \end{array} \right)$ and $\sigma_z=\left( \begin{array}{cc} 1 & 0 \\ 0 &-1 \end{array} \right)$ are the Pauli matrices, $I$ is the identity matrix, and $\vec{\sigma} = \left( \sigma_x, \sigma_y, \sigma_z \right)$ is the vector of Pauli matrices. Here, $\hat{n}=(\sin 2\theta \cos \phi, \sin 2\theta \sin \phi, \cos2\theta)$ is a unit vector expressed in spherical coordinates. By changing the values of $\theta$ and $\phi$, the unit vector $\hat{n}$ points to different points on a sphere of unit length, known as the \emph{Bloch sphere}. Note that the angle that the unit vector makes with the $z$-axis is $2\theta$. In particular, this means that orthogonal states lie on opposite ends of the Bloch sphere.

We could generalize this notion to mixed states, where the density operator generalizes to
\begin{equation}
\rho=\sum_i |\psi_i \rangle p_i \langle \psi_i |
\end{equation}
Here, $\rho$ is written in its diagonal representation, where $p_i$ are its eigenvalues and $|\psi_i\rangle$ are its eigenkets.

Since mixed states are convex combinations of pure states, we could represent each pure state in the mixture in terms of the Bloch representation.
This gives us
\begin{eqnarray}
\rho&=& \sum_i  p_i \frac{1}{2} (I+ \hat{n}_i \cdot \vec{\sigma}) \nonumber\\ &=&\frac{1}{2} \left[ I+ \left( \sum_i p_i \hat{n}_i \right) \cdot \vec{\sigma} \right] \nonumber\\ &=& \frac{1}{2} (I + \vec{n}\cdot\vec{\sigma})
\end{eqnarray}
where $\vec{n}= \sum_i p_i \hat{n}_i$ is a vector with a length less than unity\footnote{because $|\vec{n}|=|\sum_i p_i \hat{n}|\leq \sum_i p_i |n_i| = \sum_i p_i =1$, where we have used the triangle inequality.}, except in the special case when the state is pure. The vector $\vec{n}$ is known as the \emph{Bloch vector}.

The Bloch sphere representation gives us a geometrical interpretation of the purity of a state. In general, points on the surface of the Bloch sphere represent pure states, and points in the interior of the Bloch sphere represent mixed states. The completely mixed state maps to the center of the Bloch sphere.

It is now important to be able to quantify the purity of a state. Geometrically, the length of the Bloch vector $\vec{n}$ gives us a direct indication of the purity of a state, since it is a measure of the distance of the point mapped onto the Bloch sphere from the center of the sphere.  We define the purity $\mu$ of a state $\rho$ to be the square of the length of the Bloch vector, i.e. $\mu = |\vec{n}|^2$. To be able to express the purity $\mu$ of a state $\rho$ directly in terms of $\rho$, we consider
\begin{eqnarray}
\textrm{tr}(\rho^2) = \frac{1}{4} \textrm{tr}(I + 2 \vec{n} \cdot \vec{\sigma} + (\vec{n} \cdot \vec\sigma)^2) = \frac{1}{2} (1+|\vec{n}|^2)
\end{eqnarray}

Hence, the purity $\mu$ of a state $\rho$ is given by
\begin{equation}
\mu = 2 \textrm{tr}(\rho^2) -1
\end{equation}

There are, in fact, other definitions that are suited to quantify purity, for example the von Neumann entropy, which has close connections with the Boltzmann definition of entropy, or the linear entropy (which is $1-\mu)$ etc. \cite{6}. Both these definitions, however, quantify the mixedness of a state, where mixedness is $1-\mu$.

\newtheorem{Definition}{Definition}
\newtheorem{Proposition}[Definition]{Proposition}
\newtheorem{Theorem}[Definition]{Theorem}

\subsection{Reduced Density Operator \label{AppendA}} We review some properties of the reduced density operator. Suppose we have a composite system $AB$ described by the density operator $\rho_{AB}$. If we want to describe the observable quantities pertaining to subsystem $A$ without any reference to subsystem $B$, how would we do so? It turns out that the partial trace operation is the unique operation which is able to answer this question.
The partial trace over, say subsystem $B$, is defined by
\begin{equation}
\textrm{tr}_B (|a_1 \rangle \langle a_2 | \otimes |b_1 \rangle \langle b_2 |) \equiv |a_1 \rangle \textrm{tr}(|b_1 \rangle \langle b_2|)\langle a_2|
\end{equation}
where the kets $|a_i\rangle$ belong to subsystem $A$ and $|b_i \rangle$ belong to subsystem $B$, and the trace operation, in our context, which switches kets and bras, is defined\footnote{In matrix representation, this definition implies that the trace of a matrix is the sum of its diagonal elements, since $\textrm{tr} A = \textrm{tr} \sum_{ij} |i \rangle \langle i | A |j \rangle \langle j|=\sum_i \langle i |A |i \rangle$. This is seen by inserting the completeness relation on both sides of A, using trace to swop the kets and bras so that the resulting delta function kills all terms where $i \neq j$.} by
\begin{equation}
\textrm{tr}(|b_1 \rangle \langle b_2|)=  \langle b_2|b_1 \rangle
\end{equation}

With this definition\footnote{In matrix representation, if
\begin{equation} A=\left( \begin{array}{cccc} a_{0000}& a_{0001} &a_{0010} &a_{0011} \\ a_{0100}& a_{0101}&a_{0110}&a_{0111} \\ a_{1000}& a_{1001}&a_{1010}&a_{1011} \\ a_{1100}& a_{1101}&a_{1110}&a_{1111} \end{array} \right)\label{matrixrep} \end{equation} then, according to the definition, we would find\footnote{This is because $A = \sum_{klmn} |kl \rangle a_{klmn} \langle mn |$ when the Eq. (\ref{matrixrep}) is written in the basis $\{|kl \rangle \langle mn|\}$. Tracing out subsystem $B$ gives us $\textrm{tr}_B A = \sum_{km} |k \rangle \left(\sum_{ln} \langle n|l\rangle a_{klmn} \right) \langle m| = \sum_{km} |k\rangle (a_{k0m0}+a_{k1m1}) \langle m|$, which is essentially Eq. (\ref{matrixrep2}).} that \begin{equation}\textrm{tr}_B A= \left(\begin{array}{cc} a_{0000}+a_{0101} & a_{0010}+a_{0111} \\ a_{1000}+a_{1101} &a_{1010}+a_{1111} \end{array}\right)\label{matrixrep2}\end{equation}}, the observable quantities pertaining to subsystem $A$ are completely described by the reduced density operator system $A$, which is defined by
\begin{equation}
\rho_A \equiv  \textrm{tr}_B (\rho_{AB})
\end{equation}

Here, we describe\footnote{We refer the reader to Page 107 of the text `Quantum Computation and Quantum Information' by Nielsen and Chuang \cite{8}, which explains why the partial trace operation is used to describe part of a larger system.} why tracing out subsystem $B$ to obtain the reduced density operator of system $A$ describes the observable quantities pertaining to subsystem $A$: Suppose that $M$ is an observable pertaining to system $A$, which is described by the density operator $\rho_A$. Then $M\otimes I$ is the corresponding observable on the composite system $AB$ that has the density operator $\rho_{AB}$, because the measurement on $A$ is local and does not affect $B$, and so the local operation on system $B$ is the identity when $M$ is performed on $A$. Now, any measurement averages should be the same whether we compute it via $\rho_A$ or $\rho_{AB}$. Hence,
\begin{equation}
\textrm{tr}(M\rho_A) = \textrm{tr}((M\otimes I) \rho_{AB})
\end{equation}
Our task now is to express $\rho_A$ in terms of $\rho_{AB}$, i.e. we need to find the map $f$ such that $\textrm{tr}(M f(\rho_{AB})) = \textrm{tr}((M\otimes I) \rho_{AB})$, where $f(\rho_{AB}) = \rho_A$. It turns out that the unique map that satisfies this property is the partial trace operation, i.e. $f=\textrm{tr}_B$. After all, $\textrm{tr} ((M \otimes I) \rho_{AB}) = \textrm{tr}_A(M \textrm{tr}_B \rho_{AB})= \textrm{tr}(M \textrm{tr}_B \rho_{AB})$.

\section{Bell's Theorem}
In this section, we present Bell's theorem in the form derived by CHSH. Bell's Theorem may be stated as
\begin{Theorem}[Bell's theorem] Quantum mechanics is incompatible with any local hidden variable theory.
\end{Theorem}

To explain Bell's theorem, suppose that a source produces pairs of particles, say electrons, and for each pair, it sends one particle to Alice and the other to Bob, where Alice and Bob are two parties that are spacelike separated. Alice and Bob then measure some binary\footnote{binary here means that the attribute can take only one of two values.} attribute of the particles, say spin. Assume that the results of the measurements are correlated. The question that we wish to address is the following: Do the electrons have a pre-determined spin before the measurement was being made? The answer to this question is addressed by Bell's Theorem.

Suppose that there is a local hidden mechanism that determines the outcome of the measurements. Let $\lambda$ be this hidden variable, and let $\rho(\lambda)$ be the probability distribution of $\lambda$. Since $\rho(\lambda)$ is a probability distribution, it must satisfy the properties: $\rho(\lambda) \geq 0$ and $\int d\lambda \rho(\lambda) =1$. Let $\hat{a}$ and $\hat{b}$ be the directions in which Alice and Bob respectively orientate their measurement apparatus. Let $A(\hat{a},\lambda)$ and $B(\hat{b},\lambda)$ be the outcomes measured by Alice and Bob respectively. Notice that because we assume that the hidden mechanism is local, $A(\hat{a},\lambda)$ is not a function of the direction $\hat{b}$ that Bob chooses, and $B(\hat{b},\lambda)$ is not a function of the direction $\hat{a}$ that Alice chooses. This is always possible because Alice and Bob can choose the directions $\hat{a}$ and $\hat{b}$, respectively, randomly just before they make the measurement, and since we assume that the setup is local, there is no way that Alice's apparatus can \textit{know} the direction that Bob chooses, and vice versa. Because the measurement outcomes can be only either $+1$ or $-1$, hence $A(\hat{a},\lambda),  B(\hat{b},\lambda) = \pm 1$. We define the \textit{Bell correlation} $C(\hat{a},\hat{b})$ to be the expectation value of the product of the measurement outcome (either +1 or -1) of Alice and the measurement outcome of Bob (also either +1 or -1), i.e. $\langle A(\hat{a},\lambda),B(\hat{b},\lambda)\rangle$. Explicitly,
\begin{equation}
C(\hat{a},\hat{b})\equiv \int d\lambda A(\hat{a},\lambda) B(\hat{b},\lambda) \rho(\lambda)
\end{equation}

Now, suppose that Alice randomly chooses between the two directions $\hat{a}_1$ and $\hat{a}_2$ just before she makes the measurement, and Bob does the same with the directions $\hat{b}_1$ and $\hat{b}_2$. Then suppose we consider the function $F$ defined as
\begin{eqnarray}
F &=& A(\hat{a}_2,\lambda)B(\hat{b}_2,\lambda)-A(\hat{a}_1,\lambda)B(\hat{b}_1,\lambda) -A(\hat{a}_1,\lambda)B(\hat{b}_2,\lambda)-A(\hat{a}_2,\lambda)B(\hat{b}_1,\lambda) \nonumber\\ &=& A(\hat{a_2},\lambda)\left[B(\hat{b}_2,\lambda)-B(\hat{b}_1,\lambda)\right] - A(\hat{a}_1,\lambda)\left[B(\hat{b}_1,\lambda)+B(\hat{b}_2,\lambda)\right]
\end{eqnarray}
Because $A(\hat{a},\lambda)$ and $B(\hat{b},\lambda)$ can take on the values of only $1$ and $-1$ for all $\hat{a}$ and $\hat{b}$, $F$ can take on the value of only $+2$ or $-2$. This can be seen by testing all $16$ possible combinations of the values of $A(\hat{a}_1,\lambda), A(\hat{a}_2,\lambda), B(\hat{a}_1,\lambda), B(\hat{a}_2,\lambda)$. This implies that the value of $\int d \lambda \rho(\lambda) F$ is bounded by $-2$ and $2$.

Hence, we arrive at the \textit{CHSH Inequality}:
\begin{equation}
-2 \leq C(\hat{a}_2,\hat{b}_2)-C(\hat{a}_1,\hat{b}_1)-C(\hat{a}_1,\hat{b}_2)-C(\hat{a}_2,\hat{b}_1) \leq 2\label{bellchsh}\end{equation}

Our next step is to show that quantum mechanics allows correlations that violate the CHSH inequality. To do this, we use the \textit{singlet} state $|\psi_-\rangle=(|01\rangle-|10 \rangle) \frac{1}{\sqrt{2}}$. This state is in fact an example of an entangled state, but we will defer the discussion of entanglement to the next section, highlighting in this section only the fact that entangled states can have non-classical correlations.

Let us first calculate the probability $p_{kl} (\hat{a},\hat{b})$ of Alice obtaining a measurement value of $k$ as she measures her particle along the axis $\hat{a}$ and Bob obtaining a measurement value of $l$ as he measures his particle along the axis $\hat{b}$. In this case, we represent $+1$ by $0$ and $-1$ by $1$ to be in agreement with the labels of the computational basis used to describe the singlet.

Hence,
\begin{equation}
p_{kl} (\hat{a},\hat{b})=\textrm{tr} \{ |\psi_- \rangle \langle \psi_-| P_k (\hat{a}) \otimes P_l(\hat{b}) \} \label{b1}
\end{equation} where $P_k (\hat{a})$ and $P_l(\hat{b})$ are the projectors onto the respective two outcomes. Since projectors are pure states, we could represent them using the Bloch sphere representation. The two possible outcomes of each measurement are orthogonal to each other, and so are located on opposite ends of the Bloch sphere. Hence, we could write $P_k (\hat{a})=\frac{1}{2} (I+(-1)^k \hat{a} \cdot \vec{\sigma})$ and $P_l (\hat{b})=\frac{1}{2} (I+(-1)^l \hat{b} \cdot \vec{\sigma})$. Using the relation $|\psi_- \rangle \langle \psi_-|=\frac{1}{4} (I \otimes I -\sigma_x \otimes \sigma_x-\sigma_y \otimes \sigma_y-\sigma_z \otimes \sigma_z)=\frac{1}{4}(I-\vec{\sigma}^{(A)}\cdot \vec{\sigma}^{(B)})$,  Eq. (\ref{b1}) becomes
\begin{equation}
p_{kl} (\hat{a},\hat{b}) = \frac{1}{16} \textrm{tr} \left\{ (I-\vec{\sigma}^{(A)} \cdot \vec{\sigma}^{(B)}) (I+(-1)^k \hat{a} \cdot \vec{\sigma}^{(A)})(I+(-1)^l \hat{b} \cdot \vec{\sigma}^{(B)}) \right\} \label{bell2}
\end{equation} where the superscripts $A$ and $B$ indicate that the operators belong to subsystems $A$ and $B$ respectively, and so we could do away with the $\otimes$-symbol.
Because Pauli matrices are traceless, Eq. (\ref{bell2}) simplifies to
\begin{eqnarray}
p_{kl} (\hat{a},\hat{b}) &=& \frac{1}{16} \left[\textrm{tr}(I) - (-1)^{k+l} \textrm{tr}\{\vec{\sigma^{(A)}}\cdot\vec{\sigma^{(B)}} \hat{a} \cdot \vec{\sigma}^{(A)}\hat{b} \cdot \vec{\sigma}^{(B)}\}\right] \nonumber\\ &=& \frac{1}{16} \left[4-(-1)^{k+l} \textrm{tr}_A \left\{ \hat{a} \cdot \vec{\sigma}^{(A)} \textrm{tr}_B \{\vec{\sigma}^{(A)} \cdot \vec{\sigma}^{(B)} \hat{b} \cdot \vec{\sigma}^{(B)} \}\right\}\right] \nonumber\\ &=& \frac{1}{16} \left[4-2 (-1)^{k+l} \textrm{tr}_A \left\{\hat{a} \cdot \vec{\sigma}^{(A)} \hat{b} \cdot \vec{\sigma}^{(A)} \right\}\right] \nonumber\\ &=& \frac{1}{4} \left[1-(-1)^{k+l} \hat{a} \cdot \hat{b} \right]
\end{eqnarray} where we have repeatedly used the identity $\vec{a} \cdot \vec{\sigma} \ \vec{b} \cdot \vec{\sigma} = \vec{a} \cdot \vec{b} + i (\vec{a} \times \vec{b} ) \cdot \vec{\sigma}$.

Hence, the Bell correlation is given by
\begin{equation}
C(\hat{a},\hat{b}) = \sum_{k,l=0}^1 p_{kl} (\hat{a},\hat{b}) (-1)^{k+l} = \frac{1}{4} \sum_{k,l=0}^1 \left[(-1)^{k+l} - \hat{a} \cdot \hat{b} \right] = -\hat{a} \cdot \hat{b} \label{bell4}
\end{equation}

As before, Alice and Bob randomly choose between axes $a_1$, $a_2$ and $b_1$, $b_2$ respectively. Suppose that Alice and Bob agree to choose the axes such that $a_1$, $a_2$ are orthogonal to each other, and $b_1$ and $b_2$ are orthogonal to each other, such that $\hat{b}_1$ is $45^\circ$ anticlockwise of $\hat{a}_2$; $\hat{a}_1$ is $45^\circ$ anticlockwise of $\hat{b}_1$; and $\hat{b}_2$ is $45^\circ$ anticlockwise of $\hat{a}_1$, i.e. the angle between the pairs $\hat{a}_1$ and $\hat{b}_1$; $\hat{a}_1$ and $\hat{b}_2$; $\hat{a}_2$ and $\hat{b}_1$ is  $45^\circ$, while the angle subtended between $\hat{a}_2$ and $\hat{b}_2$ is $135^\circ$. By using Eq. (\ref{bell4}), we arrive at the following Bell correlations: $C(\hat{a}_1,\hat{b}_1) = C(\hat{a}_1,\hat{b}_2)= C(\hat{a}_2,\hat{b}_1) = - \cos 45^\circ = -1/\sqrt{2}$ and $C(\hat{a}_2,\hat{b}_2) = - \cos 135^\circ = 1/\sqrt{2}$.

This construction leads us to a startling conclusion: we obtain
\begin{equation}
C(\hat{a}_2,\hat{b}_2)-C(\hat{a}_1,\hat{b}_1)-C(\hat{a}_1,\hat{b}_2)-C(\hat{a}_2,\hat{b}_1) = 2\sqrt{2}\end{equation}
in direct violation of the CHSH inequality defined in Eq. (\ref{bellchsh}), because $2\sqrt{2} >2$. This tells us that the assumptions used when deriving the CHSH inequality are wrong.

Recall that the two assumptions that are used in the derivation are locality and realism. Locality enters the argument when we assume that $A(\hat{a},\lambda)$ is not a function of the direction $\hat{b}$ that Bob chooses, and $B(\hat{b},\lambda)$ is not a function of the direction $\hat{a}$ that Alice chooses. Realism enters the argument when we assume that a hidden variable $\lambda$ exists.

Bell's theorem forces us to choose between local realism and quantum mechanics. The CHSH inequality gives us a way to test if our universe obeys local realism, because the argument above shows that both local realism and quantum mechanics cannot be simultaneously true. As mentioned in the introduction, several experiments have been carried out to test the above results. Most of these experiments have been in favor of quantum mechanics. These results prompt us to study such non-classical correlations, which we -- following Schr\"{o}dinger \cite{6.3} -- call entanglement.

\section{Entanglement of bipartite systems}
According to the Stanford Encyclopedia of Philosophy \cite{7}, quantum entanglement is a physical resource, like energy, associated with the peculiar nonclassical correlations that are possible between separated quantum systems. This leads to the following definition for the entanglement of a bipartite state.

\begin{Definition}[Entanglement]  A state $\rho_{AB}$ is entangled if it cannot be prepared via local operations and classical communication (LOCC) by two parties who are spatially separated. \label{def1}
\end{Definition}

Any state that is not entangled is called \emph{separable}.
By LOCC, we mean that both parties are able to do nothing more than apply local operations on their individual subsystems or communicate via a classical channel. We first discuss the entanglement of pure bipartite states, before we show a generalization to mixed bipartite states.

\subsection{Pure bipartite states}
Definition \ref{def1} leads to the following proposition for pure bipartite states.

\begin{Proposition} A state $|\psi_{AB} \rangle$ is entangled if and only if there do not exist states $|\phi \rangle$ and $|\varphi \rangle$ such that $|\psi_{AB} \rangle = |\phi \rangle \otimes|\varphi \rangle $.
\end{Proposition}

To be able to determine whether we could express a bipartite state $|\psi_{AB} \rangle$ in terms of a tensor product $|\phi \rangle \otimes|\varphi \rangle$, we use the \textit{Schmidt decomposition}.

\begin{Theorem}[Schmidt decomposition] If $|\psi_{AB} \rangle$ is a pure state of a composite system AB, then there exists orthonormal bases $\{|\alpha_A \rangle\}$ for subsystem A, and $\{|\alpha_B\rangle\}$ for subsystem B, such that
\begin{equation}
|\psi_{AB} \rangle = \sum_\alpha |\alpha_A \rangle |\alpha_B \rangle \sqrt{p_\alpha}
\end{equation}
where $p_\alpha \geq 0$ for all $\alpha$ and $\sum_\alpha p_\alpha =1$.
\end{Theorem}
\textsl{Proof.} We first write an expression for the most general bipartite pure state.
\begin{equation}
|\psi_{AB}\rangle = \sum_{kl} |kl \rangle \psi_{kl} \label{schm1}
\end{equation}
where $|kl \rangle$ are the basis kets and $\psi_{kl}$ are complex coefficients obeying the normalization constraint. Note that when we write the basis ket of a composite system $|kl \rangle$, $k$ refers to subsystem A and $l$ to subsystem B.

We now write down the spectral decomposition of the reduced density operator $\rho_A$ for subsystem A.
\begin{equation}
\rho_A = \sum_\alpha |\alpha_A \rangle p_\alpha \langle \alpha_A| \label{schm2}
\end{equation}
where $|\alpha_A \rangle$ are the eigenkets of $\rho_A$ and $p_\alpha$ are its eigenvalues. Since the eigenvalues of the reduced density operator are probabilities, $\sum_\alpha p_\alpha =1$.

We now insert the completeness relation into Eq. (\ref{schm1}). This gives us
\begin{eqnarray}
|\psi_{AB}\rangle &=& \sum_{kl} \left( \sum_\alpha |\alpha_A \rangle \langle \alpha_A|\otimes I \right) |kl \rangle \psi_{kl} \label{schm3}  \\&=& \sum_\alpha |\alpha_A \rangle \left(\sum_{kl} |l \rangle \langle \alpha_A|k \rangle \psi_{kl} \right) \label{schm4}
\end{eqnarray}

Let us denote the un-normalized state $\sum_{kl} |l \rangle \langle \alpha_A|k \rangle \psi_{kl}$ in the parenthesis in Eq. (\ref{schm4}) by $|\tilde{\alpha}_B \rangle$, and proceed to find its norm.
\begin{eqnarray}
\langle \tilde{\alpha}_B | \tilde{\alpha}_B \rangle &=& \sum_{kk'll'} \psi_{k'l'}^* \langle k' |\alpha_A \rangle \langle l'|l \rangle \langle \alpha_A |k \rangle \psi_{kl} \nonumber\\ &=& \langle \alpha_A | \left( \sum_{kk'l} |k \rangle \psi_{k'l}^* \psi_{kl} \langle k'| \right) |\alpha_A \rangle \label{schm5}
\end{eqnarray}

But
\begin{eqnarray}
\rho_A = \textrm{tr}_B (\rho_{AB}) = \textrm{tr}_B \left(\sum_{kk'll'} |kl \rangle \psi_{kl} \psi_{k'l'}^* \langle k'l'|\right) =\sum_{kk'l} |k \rangle \psi_{k'l}^* \psi_{kl} \langle k'| \label{schm6}
\end{eqnarray}

Comparing Eq. (\ref{schm5}) and Eq. (\ref{schm6}) implies that the square of the norm of $| \tilde{\alpha}_B \rangle$ is
\begin{eqnarray}
\langle \tilde{\alpha}_B | \tilde{\alpha}_B \rangle = \langle \alpha_A| \rho_A |\alpha_A \rangle = p_\alpha
\end{eqnarray}
since $|\alpha_A \rangle$ is an eigenket of $\rho_A$.

Hence, we define $|\alpha_B \rangle = |\tilde{\alpha}_B \rangle \frac{1}{\sqrt{p_\alpha}}$ to be the normalized kets belonging to subsystem B. We, thus, arrive at the Schmidt decomposition $|\psi_{AB} \rangle = \sum_\alpha |\alpha_A \rangle |\alpha_B \rangle \sqrt{p_\alpha}$. By symmetry, $|\alpha_B \rangle$ are the eigenkets of the reduced density operator $\rho_B$, with $\sqrt{p_\alpha}$ as the eigenvalues. Since the reduced density operators are Hermitian, we could always find orthonormal bases  $\{|\alpha_A \rangle\}$ for subsystem A, and $\{|\alpha_B\rangle\}$ for subsystem B. QED.

Here, the bases $|\alpha_A \rangle$ and $|\alpha_B \rangle$ are known as the \textit{Schmidt bases} for A and B respectively, $p_\alpha$ are the \textit{Schmidt coefficients}, and the number of nonzero values of $p_\alpha$ is known as the \textit{Schmidt number}.

What we have essentially done is to write any arbitrary state in terms of an orthonormal basis that are eigenkets of the respective reduced density operators, for which the eigenvalues for both reduced density operators are the same. This representation is important because it allows us to see immediately whether a state is entangled or separable. This is explained in the following proposition.

\begin{Proposition} A pure state $|\psi_{AB} \rangle$ is entangled if and only if its Schmidt number is greater than unity. \label{thm4}
\end{Proposition}
\textsl{Proof.}
If the Schmidt number of a state $\psi_{AB}$ is equal to unity, then $\psi_{AB}= |\alpha_A \rangle |\alpha_B \rangle$. Hence, by definition, $\psi_{AB}$ is separable. The converse is also true. QED.

We are led now to the following theorem.
\begin{Theorem} A pure state $|\psi_{AB}\rangle$ is entangled if and only if $\rho_A$ (or $\rho_B$) is not pure.
\end{Theorem}
\textsl{Proof. }The density operator $\rho_{AB}$ is given by
\begin{equation} \rho_{AB}= |\psi_{AB} \rangle \langle \psi_{AB}| = \sum_{\alpha \alpha'} |\alpha_A \rangle |\alpha_B \rangle \sqrt{p_\alpha p_{\alpha '}} \langle \alpha| \langle \alpha'|
\end{equation}

Hence,
\begin{equation} \rho_B = \textrm{tr}_A (\rho_{AB}) = \sum_\alpha |\alpha_B \rangle p_\alpha \langle \alpha_B|
\end{equation}

Similarly,
\begin{equation} \rho_A = \sum_\alpha |\alpha_A \rangle p_\alpha \langle \alpha_A|
\end{equation}

From Proposition \ref{thm4}, the reduced density matrix is pure if and only if the Schmidt number is equal to one. QED.

In summary, the Schmidt decomposition is a test for entanglement of a pure bipartite state. Given an arbitrary pure state $|\psi \rangle$, the recipe is to the first write $|\psi \rangle$ in the Schmidt decomposition, and then determine the Schmidt number, which tells us if a state is entangled or separable.

\subsection{Mixed States}
The entanglement of mixed states is not completely understood. The reason for the problem with mixed states aries because the quantum content of the correlations is hidden behind the classical correlations in a mixed state \cite{9}. However, for systems where the dimensions of the subsystems are $2$ and $2$ or $3$ respectively, we could find a necessary and sufficient condition for a state to be entangled.

Before we describe this, we first use Definition \ref{def1} to arrive at the following proposition for mixed bipartite states.

\begin{Proposition} A mixed state $\rho_{AB}$ is entangled if and only if there do not exist states $\rho_A$ and $\rho_B$ such that $\rho_{AB}$ can be written\footnote{Notice that only one index of summation is used, since using two is equivalent to using one.} as $\rho_{AB} = \sum_k p_k \rho_A^{(k)} \otimes \rho_B^{(k)}$, where $\sum_k p_k=1$. \label{prop2}
\end{Proposition}
We could justify this proposition by arguing that LOCC allows two spacelike separated subsystems A and B to do nothing more than create mixed states in their own isolated subsystem and communicate via a classical channel to mix them according to certain agreed probabilities. Hence, $\rho_{AB} = \sum_k p_k \rho_A^{(k)} \otimes \rho_B^{(k)}$ is the most general separable state that two spacelike separated parties can create via LOCC.

This leads us to the Peres-Horodecki criterion, which gives a necessary condition for a bipartite mixed state to be separable. The theorem however requires us to define the partial transposition map and what the statement $A\geq 0$ means, where A is an operator, and so we will review these first. The \textit{partial transposition} map $\rho_{AB}^{T_B}$ of a bipartite density operator $\rho_{AB}$ is defined as follows: If
\begin{equation}
 \rho_{AB} = \sum_{klmn}|kl\rangle \rho_{klmn} \langle mn|
\end{equation}
then
\begin{equation}
\rho_{AB}^{T_B} = \sum_{klmn}\rho_{klmn} |k\rangle \langle m| \otimes (|l\rangle \langle n|)^T
\end{equation}
where T is the usual matrix transposition.

We use the shorthand $A\geq 0$, for an operator $A$, to mean that $A$ is positive semidefinite. A matrix $A$ is positive semidefinite, if for all $|\psi\rangle$, $\langle \psi| A |\psi \rangle$ is nonnegative. It may be shown that every positive semidefinite matrix is Hermitian, and hence diagonalizable, and that the definition leads to the statement: A matrix A is positive semidefinite if and only if all its eigenvalues are non-negative.

We now state and proof the Peres-Horodecki criterion.
\begin{Theorem}[Peres-Horodecki criterion] If a mixed state $\rho_{AB}$ is separable, then $\rho_{AB}^{T_B} \geq 0$.
\end{Theorem}
\textsl{Proof.} If $\rho_{AB}$ is separable, then we could write $\rho_{AB} = \sum_k p_k \rho_A^{(k)} \otimes \rho_B^{(k)}$, by Proposition (\ref{prop2}). Hence, by definition of partial transposition, $\rho_{AB}^{T_B} = \sum_k p_k \rho_A^{(k)} \otimes (\rho_B^{(k)})^T$. But the transpose of a density operator is also a density operator. To see why this is true, we recall the conditions that characterize a density operator. An operator $A$ is a density operator if and only if $\textrm{tr}(A) =1$ and $A\geq0$. But the trace and eigenvalues of a matrix remain the same under transposition. Hence, $(\rho_B^{(k)})^T$ is also a density operator. Since the tensor product of two density operators is also a density operator, hence $\rho_{AB}^{T_B}$ is also a density operator. This implies that $\rho_{AB}^{T_B}\geq 0$ QED.

It turns out that the converse of the Peres-Horodecki criterion is true only in special cases. This was proven by M. Horodecki, P. Horodecki, and R. Horodecki \cite{10}. We state the theorem without proof in these notes.

\begin{Theorem} A mixed state $\rho_{AB}$ with $dim(A) =2$ and $dim(B)=2$ or $3$ is separable if and only if $\rho_{AB}^{T_B} \geq 0$.
\end{Theorem}

This theorem allows us to distinguish between separable and mixed states for systems that satisfy the conditions stated in the theorem, by looking at the positive-semidefiniteness of the partial transpose of the density operator.

\section{Quantum teleportation: an application of entanglement \label{telep}}
In the past decade, several experiments have been carried out to test teleportation with photons \cite{11,12} and atoms \cite{13,14}. In this section, we describe a simple example of how a single pure state qubit may be teleported from one location to another. The problem can be described as follows: Suppose that Alice has in her possession a pure state, given by $|\varphi \rangle= |0\rangle \alpha + |1\rangle \beta$, where $|\alpha|^2+|\beta|^2=1$, and she wants to send it to Bob. Let us suppose that Alice may not know\footnote{Even if Alice knows the values of $\alpha$ and $\beta$, she would require an infinite number of bits to completely describe the state to Bob because  $\alpha$ and $\beta$ may be irrational.} the values of the complex numbers $\alpha$ and $\beta$ in general, and so it is impossible for her to describe the qubit to Bob in order that Bob may reconstruct the qubit in his laboratory, i.e. it is impossible for Alice to send the pure state to Bob using classical communication. However, we could solve the problem if Alice and Bob resort to using teleportation.

Before we describe the process of teleportation, let us define a basis for a 2-qubit state such that the basis elements are all maximally entangled states. Here, a maximally entangled 2-qubit state is a state in which the reduced density operator of either of the subsystems is maximally mixed. The orthonormal basis $\{|\phi_{+} \rangle, |\phi_{-} \rangle, |\psi_{+} \rangle, |\phi_{-} \rangle\}$ where
\begin{eqnarray}
|\phi_{\pm} \rangle = (|00 \rangle \pm |11 \rangle ) \frac{1}{\sqrt{2}} \nonumber \\ |\psi_{\pm}\rangle = (|01 \rangle \pm |10 \rangle ) \frac{1}{\sqrt{2}} \label{bellbasis}
\end{eqnarray} satisfies this property. The elements in this basis are known as \textit{Bell states}. We could check that tracing out any of the subsystems will give a reduced density matrix that is equal to $\frac{1}{2} I$. This implies that the Bell states are maximally entangled.

We will now describe the teleportation protocol. Let us place between Alice and Bob a source which produces a maximally entangled state, $|\phi_+\rangle = (|00 \rangle +|11 \rangle)\frac{1}{\sqrt{2}}$, so that it sends the first particle to Alice and the second particle to Bob.

The composite system of all 3 particles (the qubit that Alice wants to teleport as well as the two particles from the source) can be written as
\begin{eqnarray}
|\varphi \rangle |\phi_+ \rangle &=& (|0 \rangle \alpha + |1 \rangle \beta)(|00 \rangle +|11 \rangle) \frac{1}{\sqrt{2}} \nonumber \\&=& |000 \rangle \frac{\alpha}{\sqrt{2}}+|011 \rangle \frac{\alpha}{\sqrt{2}}+|100 \rangle \frac{\beta}{\sqrt{2}}+|111 \rangle \frac{\beta}{\sqrt{2}} \label{tele1}
\end{eqnarray}
Note that in the notation $|a_1 a_2 a_3\rangle$, the first two particles described by the labels $a_1$ and $a_2$ refer to Alice's two particles (the qubit she originally had as well one received from the source) and the third particle described by the label $a_3$ refers to the particle that Bob received from the source.

Now we could express the two qubits possessed by Alice and Bob in terms of the Bell basis rather than the computational basis. We use Eq. (\ref{bellbasis}) to arrive at
\begin{eqnarray}
|00 \rangle = (|\phi_+ \rangle + |\phi_- \rangle) \frac{1}{\sqrt{2}} \qquad \qquad
|11 \rangle = (|\phi_+ \rangle - |\phi_- \rangle) \frac{1}{\sqrt{2}} \nonumber\\
|01 \rangle = (|\psi_+ \rangle + |\psi_- \rangle) \frac{1}{\sqrt{2}} \qquad \qquad
|10 \rangle = (|\psi_+ \rangle - |\psi_- \rangle) \frac{1}{\sqrt{2}} \label{bellbasis2}
\end{eqnarray}

Inserting Eq. (\ref{bellbasis2}) into Eq. (\ref{tele1}),
\begin{eqnarray}
|\varphi \rangle |\phi_+\rangle &=& (|\phi_+ \rangle + |\phi_-\rangle) |0\rangle \frac{\alpha}{\sqrt{2}} +(|\psi_+ \rangle + |\psi_-\rangle) |1\rangle \frac{\alpha}{\sqrt{2}} \nonumber\\&&+(|\psi_+ \rangle - |\psi_-\rangle) |0\rangle \frac{\beta}{\sqrt{2}}+(|\phi_+ \rangle - |\phi_-\rangle) |1\rangle \frac{\alpha}{\sqrt{2}} \nonumber\\ &=& |\phi_+ \rangle (|0 \rangle \alpha + |1 \rangle \beta) \frac{1}{2} + |\phi_- \rangle (|0 \rangle \alpha - |1 \rangle \beta) \frac{1}{2} \nonumber\\ &&+|\psi_+ \rangle (|1\rangle \alpha + |0 \rangle \beta) \frac{1}{2} + |\psi_- \rangle (|1 \rangle \alpha - |0 \rangle \beta) \frac{1}{2} \nonumber\\ &=& |\phi_+\rangle (I |\varphi\rangle) \frac{1}{2}+|\phi_-\rangle (\sigma_z |\varphi\rangle) \frac{1}{2} + |\psi_+\rangle (\sigma_x |\varphi\rangle) \frac{1}{2}+ |\psi_-\rangle (-i\sigma_y |\varphi\rangle) \frac{1}{2}
\end{eqnarray}

This tells us that the composite system is in a equal superposition of the 4 states $|\phi_+\rangle\otimes I |\varphi\rangle$, $|\phi_-\rangle \otimes\sigma_z |\varphi\rangle$, $|\psi_+\rangle \otimes\sigma_x |\varphi\rangle$ and $|\psi_-\rangle \otimes \sigma_y |\varphi\rangle$, where the probability associated with each outcome is $1/4$.

At this point, Alice will carry out a projective measurement on her 2 particles, using the projection\footnote{also known as von Neumann measurement operators} operators $|\phi_+\rangle \langle \phi_+|$, $|\phi_-\rangle \langle \phi_-|$, $|\psi_+\rangle \langle \psi_+|$ and $|\psi_-\rangle \langle \psi_-|$. This means that she will obtain one of the four bell states with a probability of 1/4. At this point, Bob's state is completely mixed (because tracing out Alice's system will give $\frac{1}{2} I$), as seen from the fact that Bob does not know the result of Alice's measurement. Alice will now need to communicate via a classical channel to tell Bob the result of her measurement. When Bob learns the result of Alice's measurement, he would then have to perform the relevant operation to recover the state that Alice wanted to teleport to him. For example, if Alice measured $|\phi_+\rangle$, then Bob will do nothing to his state; if Alice measured $|\phi_-\rangle$, Bob will act $\sigma_z$ on his qubit to recover $|\varphi\rangle$, and so on.

Following the above protocol will result in the state $\varphi=|0\rangle\alpha + |1\rangle \beta$ being teleported from Alice to Bob. Since $\alpha$ and $\beta$ are arbitrary, it means that any single qubit state can be teleproted. We justify the name teleportation given to such a process, because the state is transferred from Alice to Bob without traversing the space that separates them. Note that teleportation does not involve a transfer of matter or mass from one location to another. In the protocol described, only the structure of matter/mass is transferred from Alice to Bob. Hence, teleportation does not violate casuality, because the transfer of information does not occur faster than the speed of light, because the recreation of the qubit at Bob's location cannot occur until Alice sends Bob information about her measurement. Such information travels through a classical channel, where signals do not exceed the speed of light.

Also, note that in the above protocol, only $\log_2 4 =2$ classical bits, used to encode information about the $4$ possible outcomes of measurements, need to be sent to Bob in order for him to recover the state $|\varphi\rangle=|0\rangle\alpha + |1\rangle \beta$. Teleportation offers a significant advantage over classical communication because we are able to use just 2 bits to send information about a qubit, which would require, as we have argued, an infinite amount of classical information to characterize completely.

\section{Concluding remarks}
We have shown how entanglement was first detected by mathematical considerations, and how the CHSH inequality gives us a way to test if our universe obeyed local realism. We have shown some methods that can allow us to determine if a bipartite state is entangled or separable. Finally, we demonstrated how entanglement can be used in quantum teleportation. The importance of understanding entanglement can be overemphasized, because doing so will enable us to not only gain a more complete understanding of the true nature of our universe, but also because of the many applications that entanglement offers, of which teleportation is but one example.

\section{Acknowledgments}
I wish to thank L. C. Kwek and S. G. Tan for the discussions on quantum information and entanglement, and D. Kaszlikowski for the lectures on quantum information.

\end{document}